\documentclass[conference]{IEEEtran}
\IEEEoverridecommandlockouts
\usepackage{cite}
\usepackage{amsmath,amssymb,amsfonts}
\let\oldnl\nl
\newcommand\nonl{%
	\renewcommand{\nl}{\let\nl\oldnl}}

\usepackage{algorithm} 
\usepackage[noend]{algpseudocode}

\usepackage{graphicx}
\usepackage{textcomp}
\usepackage{xcolor}

\usepackage{tikz}
\usetikzlibrary{positioning}
\usepackage{pgfplots}
\usepackage{microtype}
\usepackage{caption,subcaption}

\usepackage{booktabs} 
\usepackage{float}
\usepackage{graphicx}

\usepackage{multirow}
\usepackage{siunitx}

\usepackage[english]{babel}
\usepackage[utf8]{inputenc}
\usepackage{calligra}

\usepackage{graphicx}
\usepackage[T1]{fontenc}

\usepackage{subcaption}
\newtheorem{theorem}{Theorem}[section]

\newtheorem{lemma}[theorem]{Lemma}

\newcommand{\eps}{\epsilon}

\def\BibTeX{{\rm B\kern-.05em{\sc i\kern-.025em b}\kern-.08em
    T\kern-.1667em\lower.7ex\hbox{E}\kern-.125emX}}
\begin{document}

\title{Detecting DDoS Attack on SDN Due to Vulnerabilities in OpenFlow}

\author{
	\IEEEauthorblockN{Sarwan Ali\IEEEauthorrefmark{1}, Maria Khalid Alvi\IEEEauthorrefmark{1}, Safi Faizullah\IEEEauthorrefmark{2}, Muhammad Asad Khan\IEEEauthorrefmark{4}, Abdullah Alshanqiti\IEEEauthorrefmark{3}, Imdadullah Khan\IEEEauthorrefmark{1}
	}
	\IEEEauthorblockA{\IEEEauthorrefmark{1}\textit{Department of Computer Science, Lahore University of Management Sciences (LUMS),} \\ Lahore, Pakistan
		\\ \IEEEauthorrefmark{1}\{16030030,maria.alvi,imdad.khan\}@lums.edu.pk}
	\IEEEauthorblockA{\IEEEauthorrefmark{2}\IEEEauthorrefmark{3}\textit{Department of Computer Science, Islamic University,} \\ Madinah, Saudi Arabia
		\\ \IEEEauthorrefmark{2} safi@iu.edu.sa, \IEEEauthorrefmark{3}amma@iu.edu.sa}
	\IEEEauthorblockA{\IEEEauthorrefmark{4}\textit{Department of Telecommunication, Hazara University,} \\ Mansehra, Pakistan
	\\ \IEEEauthorrefmark{4} asadkhan@hu.edu.pk}	
}


\maketitle

\begin{abstract}
	Software Defined Networking (\textsc{SDN}) is a network paradigm shift that facilitates comprehensive network programmability to cope with emerging new technologies such as cloud computing and big data. \textsc{SDN} facilitates simplified and centralized network management enabling it to operate in dynamic scenarios. Further, \textsc{SDN} uses the OpenFlow protocol for communication between the controller and its switches. The OpenFlow creates vulnerabilities for network attacks especially Distributed Denial of Service (DDoS).  DDoS attacks are launched from the compromised hosts connected to the \textsc{SDN} switches. In this paper, we introduce a time- and space-efficient solution for the identification of these compromised hosts. Our solution consumes less computational resources and space and does not require any special equipment.
\end{abstract}

\section{Introduction}
With the proliferation of mobile devices, and the emergence of new technologies such as the Internet of Things, and Cloud computing etc., the number of internet devices has been increased at an unprecedented rate. This has triggered a significant increase in the growth and complexity of networks on a large scale which comes with its challenges. The existing network technologies and infrastructure do not offer a scalable and easy to manage solution for such large and complex networks \cite{bannour2018_b8}. 

Software Defined Networking (\textsc{SDN}) \cite{bizanis2016_b15} stands out as an important solution among all the other proposed approaches in terms of coping with the inherent challenges of a complex and large network. 
One of the most salient feature of \textsc{SDN} is division of the data plane and the control plane. The control plane makes the decisions where to send packets, and the data plane implements these decisions and actually forwards the packets.
Additionally, optimal network operations can be achieved by utilizing the benefits of centralization i.e. making decisions with a complete view of overall network conditions from a single and centralized point. Though \textsc{SDN} offers flexibility and efficient management at a low cost, it also introduces new vulnerabilities \cite{Faizullah2018_b7}. 

OpenFlow is a standard protocol that is used for communication between switches and central controller in \textsc{SDN}. Being a relatively new paradigm, OpenFlow still has some vulnerabilities which can be exploited in many ways that directly affect the security of the network. OpenFlow deals with the incoming data packet according to the matching flow entry in the flow table of the switch. The packet may be forwarded to the controller for further processing in case of a table miss in the flow table.  This gives room to the attackers to devise an attack called Distributed Denial of Service (DDoS) attack. The malicious attacker can generate a large number of packets with missing entries in the flow table which are sent to the central controller by OpenFlow, ultimately exhausting the central controller and failing the performance of the network. Many recent studies have identified that the \textsc{SDN} paradigm is prone to DDoS attacks by malicious users \cite{hameed2018_b5} \cite{wang2019_b6}.
In this work, we are focusing on the detection of compromised hosts (attacker). As a switch may be serving many hosts, it is very important for DDoS mitigation that compromised hosts should be detected efficiently (both in terms of time and space).

Our main contributions are the following:
\begin{itemize}
	\item We developed an accurate method to study the communication patterns of each host and detect compromised (attacker/zombie) hosts.
	\item Our method is efficient in terms of time and space complexity.
\end{itemize}

The rest of this paper is organized as follows: Section \ref{background_and_problem_formulation} presents \textsc{SDN} background, DDoS attack, and problem formulation. We provide a brief review of existing methods in Section \ref{section_related_work}. Section \ref{proposed_approach} elaborates our proposed solution and in Section \ref{section_experimental_setup} we present the experimental setup and results of the proposed method. Finally, the paper is concluded in Section \ref{conclusion}. 

\section{Background and Problem Formulation}\label{background_and_problem_formulation}
In this section, we discuss the \textsc{SDN} architecture and the DDoS problem in OpenFlow based \textsc{SDN}. We begin with a brief description of \textsc{SDN} fundamentals, proceed to OpenFlow operation, DDoS attack and its consequences on the network performance. Finally, problem is formulated for finding an optimal solution.

\subsection{\textsc{SDN} Architecture}
\textsc{SDN} is one of the emerging networking technologies and one of the most important block in softwarization of communication networks. Moreover, \textsc{SDN} enables the network to be dynamic, easy to manage, flexible and scalable to adapt itself according to the given requirements. For this purpose, \textsc{SDN} separates the control plane from the data/forwarding plane and concentrate all the computing capabilities at the centralized controller. As the controller has an overall picture of the network, a better view of network bottlenecks, and centralized information processing capability, it is in a much better position to make a well-calculated decision considering the whole scenario. 

\begin{figure}[h!]
	\centering
	\includegraphics[scale=0.6,page=1]{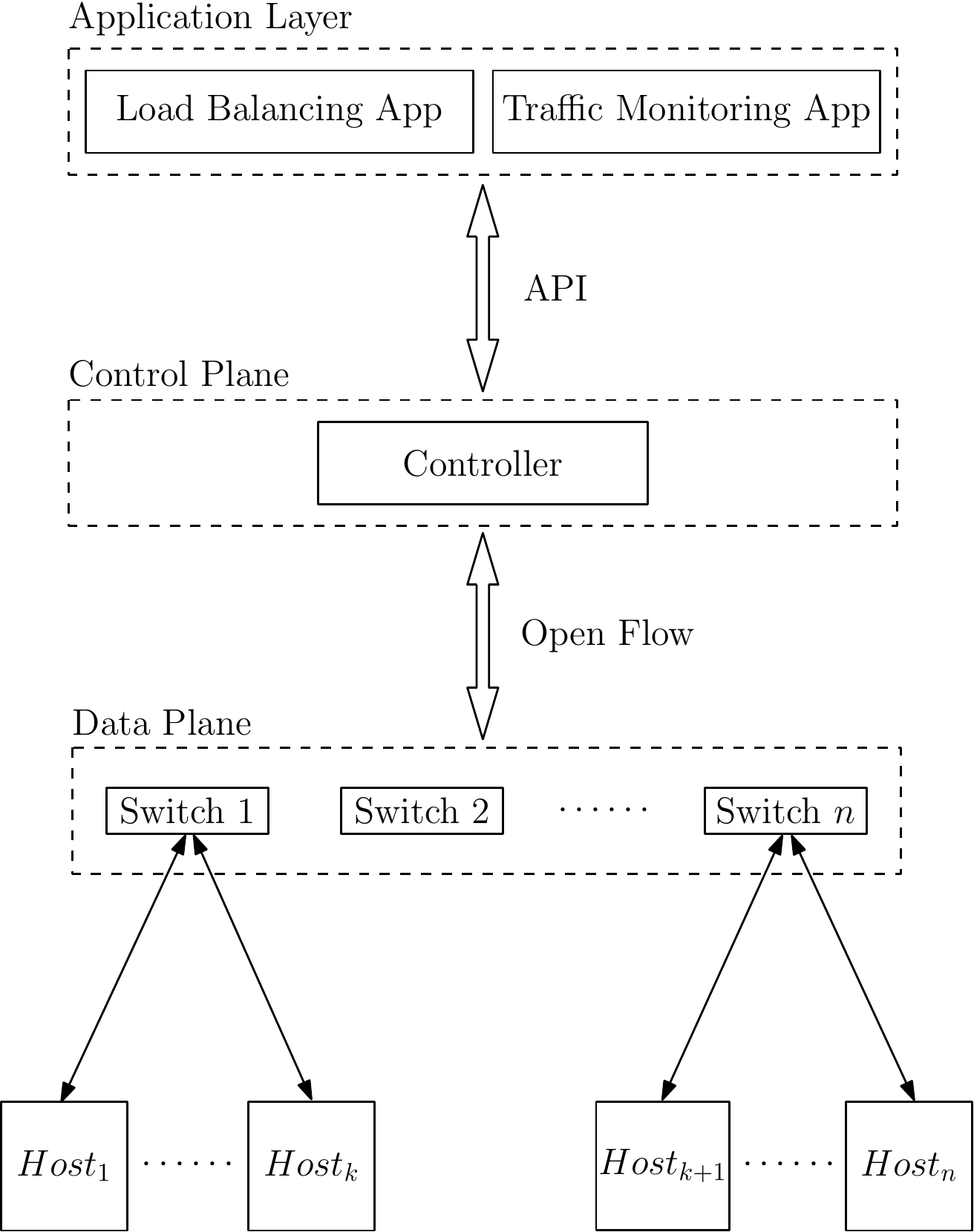}
	\caption{\textsc{SDN} Architecture.}
	\label{fig:_SDN Architecture}
\end{figure}

An \textsc{SDN} architecture consists of three layers: the First layer (application layer) is composed of applications that are responsible for the management and security of the underlying network. The second layer (control layer) provides a platform (centralized controller) for the application layer to control and manage the underlay infrastructure layer. The centralized controller act as the brain of network, performing processing and instructing switches for their operation. \textsc{SDN} controller can be seen as the operating system of a network, it facilitates automated network management and simplifies the integration of new business applications. 
Finally, the Last layer consists of physical network entities that are controlled and managed according to the application specifications and requirements via the \textsc{SDN} controller. The block diagram of \textsc{SDN} architecture is given in Figure \ref{fig:_SDN Architecture}.

\subsection{OpenFlow Protocol}
OpenFlow is a Layer $2$ protocol and closely associated with the \textsc{SDN} for communication between different \textsc{SDN} nodes specifically, the controller and switches. OpenFlow protocol enables the \textsc{SDN} controller to communicate and instruct its connected switches on how to deal with different types of incoming packets. Precisely, the \textsc{SDN} switches work as forwarding devices following certain rules which are defined by the controller\cite{li2018_b23}.
An OpenFlow enabled switch may perform the following functions in the response of incoming packets. 
\begin{itemize}
	\item Identification and categorization of incoming/ingress packets based on packet header fields. 
	\item Process the packets e.g. packet header modification. In some cases, the action in a flow entry asks for modification in the packet header e.g. change time to live, flow priority.
	\item Drop or forward the packets to a specific outgoing/egress port or \textsc{SDN} controller.
\end{itemize}
The instructions transmitted from an \textsc{SDN} Controller to an \textsc{SDN} switch are defined as \textit{flows}. Each individual flow contains different information such as match fields, flow priority, counters, actions, and flow timeouts, etc. The flows are arranged in flow tables. A switch should have at least one flow table but have the option of multiple flows tables for flexibility.  If the header of an incoming packet matches with a flow table entry, the action specified for this entry is performed, otherwise (in case of unique/non-repetitive random header) the packet is encapsulated in a message (query) and send to the controller using OpenFlow protocol. In response to a switch query, the controller processes the header of the given packet and decides either its path to forward it or drop it. Further, controller instruct switches about the specified action for any future packets of this kind.

\subsection{DDoS on \textsc{SDN} Controller}
Consequently, the centralization of processing capabilities of a network introduces the problem of a \textit{single point of failure}. In other words, if somehow the \textsc{SDN} controller is compromised or made unavailable for the switches queries, the whole network will lose its function. One of the famous attacks on the centralized controller is DDoS, where a huge number of packets with non-repetitive headers are injected from a group of compromised (zombies) hosts into the network using techniques like spoofing etc.\cite{alshamrani2017_b17} \cite {yang2018_b18}. As these packets have non-repetitive headers that result in table misses and packets are forwarded to the controller for processing. Every controller has a limited processing capacity and this flooding of unspecified packets increases processing load on it. Eventually, the load of these packets increases so much that the controller becomes unavailable for the normal network processing. A failed controller is the basic aim of DDoS attack as this results in the loss of \textsc{SDN} architecture\cite{dong2016_b2}\cite{dayal2016_b13}.

\subsection{Problem Formulation}
Let consider a \textsc{SDN} network, consists of a centralized controller ${\cal C}$ which manages a set of $N$ switches $S = \{ {s_1},{s_2},\ldots,{s_N}\}$. Further, each switch is providing services to a set of hosts/clients that are connected to it. For example switch $s_i$ is serving a set of hosts $H_{pi}=\{h_{p1}, h_{p2},\ldots,h_{pi}\}$. Moreover, the controller ${\cal C}$ with the computation capabilities of $c$ that represents the amount/number of queries that a controller can handle at a given time. Similarly, each switches $s$ has a buffer with capacity $k$ that is the amount of storage to temporary store non-repetitive packets. Further, each switch receives incoming traffic represented by $T_i$, which consists of repetitive $t_{ri}$ and non-repetitive $t_{mi}$ packets ($T_i=t_{ri} +t_{mi}$). The non-repetitive packets cause table misses and eventually forwarded to the controller for processing. Therefore, the load $L_c$ on a centralized controller will be the amount of non-repetitive packets from all the connected switches and given as follows
\begin{equation}
{L_c} = \sum_{i=1}^{N} t_{mi}
\end{equation}
Further, a DDoS attack introduces an artificial flood of non-repetitive packets on the controller \cite{xu2016_b22}. We can represent the computation capacity of a controller as the capacity to process the amount of non-repetitive packets per unit time, given as 
${C_c} = x{t_m}$, where $x$ is the maximum number of packets that the controller can process in a unit time.  
In the given scenario, we can say that a DDoS attack is successful if the number of non-repetitive packets in the controller is equal to or greater than the controller capacity which is given as follows
\begin{equation}
{C_c} \le  \sum_{i=1}^{N} t_{mi}
\end{equation}
where $t_{mi}$ denotes the number of non-repetitive packets at switch $s_i$. 

\section{Related Work}\label{section_related_work}
To cope with the increasing rate of internet devices and new internet technologies, \textsc{SDN} is considered a vital solution among all the other existing methods. Its decoupled architecture allows more flexibility, management, and low cost than the traditional networks \cite{bizanis2016_b15}. 

\textsc{SDN} uses Open Flow as a communication protocol that maintains all the communication between the switches and the main controller. Though \textsc{SDN} offers a lot of benefits over the traditional network, its architecture still needs improvement to make it more secure. Open Flow also introduces vulnerabilities in \textsc{SDN} architecture which can attract a DDoS attack \cite{dharma2015_b12}.

Many studies have been done to detect DDoS attack on \textsc{SDN} due to vulnerabilities in Open Flow \cite{bizanis2016_b15} \cite{hameed2018_b5} \cite{dayal2016_b13} \cite{bawany2017_b14}  \cite{kalkan2017_b1}. Kotani et al.  \cite{kotani2014_b9} presents a solution to filter out Packet-In messages to be processed by the centralized controller to save it from getting overburdened. Their solution suggests applying restrictions on important and less important messages based on the value in headers. The switches store the header values of the packets to be forwarded to the controller for time $t$. If the packet with existing header values arrives in $t$, it does not get forwarded to the controller in that time window. This solution does not cater for DDoS attack with non-repetitive random header patterns.  
Mousavi et al. \cite{mousavi2015_b10} introduce the idea of using entropy to identify the decrease in randomness in the flow of packets towards controllers and detect an early DDoS attack. The frequencies for the destination IP addresses are stored for window size $t$ and entropy is calculated to identify the DDoS attack when the arrived packet number reaches the window size. 

In \cite{buragohain2016_b3}, a DDoS detection framework for SDN has been proposed specifically for data centers using bounds on the rate and duration of the flow defined by analyzing the statistics collected from flows at switches. If the traffic does not comply with the defined bounds of legitimate traffic, it initiates mitigation action.

Wang et al. \cite{wang2019_b6} proposed a two-modules based scheme. The first module is to detect the anomaly in traffic by monitoring the traffic flows of switches based on rate features and asymmetric features. The second module mitigates the DDoS attack by activating the slave controller which shares the traffic load of the master controller while the master controller sends messages to switches to drop the malicious messages based on traffic analysis results. 

Dridi et al. \cite{dridi2016_b4} give a solution for DDoS attack in \textsc{SDN} by dynamically rerouting the malicious traffic, adjusting the flow timeouts and aggregating the flow rules. Lim et al. \cite{lim2015_b19} suggest to modifying the model of the controller by setting up logical queues (separate queues for each switch), and the controller should serve these logical queues with a scheduling discipline.

Most of the work that has been done previously in an attempt to detect and mitigate the problem involves data storage and analysis which consumes memory, requires complex computations, and presents the risk of false-positive and false-negative \cite{macedo2016_b21}. This raises the need for an efficient solution that consumes less memory and provides more assurance. 

\section{Proposed Approach}\label{proposed_approach}
In this section, we present the method for detecting compromised hosts. 
Consider an \textsc{SDN} architecture with a set ${\cal S} = \{s_1,s_2,\ldots,s_N\}$ of $N$ switches and that the switch $s_i \in {\cal S}$ serves $p_i$ hosts. Let ${\cal H}$ be the set of all possible unique headers that a packet can contain. 

The defining characteristic of a compromised host is that a majority of the packets received from it at the serving switch have unique (non-repetitive) headers. This forces the switch to seek forwarding instructions from the controller, thus overwhelming the controller and resulting in a DDoS attack. Each switch can (in theory) analyze patterns in the packets coming from each of its hosts and identify an attacking host. 

To undertake such an analysis, the switch will require to store the information about ingress packets of each of its hosts and determine the number of unique headers and that of repetitive headers. Formally, let $h$ be a host served by switch $s_i$ and $F$ be a $|{\cal H}|$-dimensional frequency vector, i.e. $F[j]$ is the number of packets containing header $j$ within an observation window (frequency of header $j$). 

The host $h$ can then be pinpointed as a (potentially) compromised if the majority of the non-zero frequencies in the vector $F$ are $1$'s or small positive numbers. This is so because if say $490$ of the last $500$ packets incoming from $h$ are unique, this will cause an increased load on the controller. 

The observation window could be time-based (e.g. the last few timestamps) or based on traffic volume (e.g. the latest fixed number of packets).  We take the observation window to be traffic volume-based, i.e. we assume $F$ is the headers frequency vector of the last $M$ incoming packets from the host, where $M$ is window size (total number of incoming packets).

Various statistics of the $F$ vector can be used to highlight a potentially erratic behavior. Let 
\begin{equation}
F_0 = \sum_{j\in {\cal H}} (F[i])^0
\end{equation} 
where $0^0:=0$. Thus $F_0$ is the number of unique headers used by a host. As discussed above a high value of $F_0$ for a host is a clear indication of a host being compromised. For instance, a host can be declared compromised if $F_0 / M \geq 80\%$.

We define $F_1$ and $F_2$ as following

\begin{equation}
F_1 = \sum_{j\in {\cal H}} F[i]
\end{equation}
\begin{equation}
F_2 = \sum_{j\in {\cal H}} (F[i])^2
\end{equation}

Clearly, $F_1 = M$, as it is just counting the total number of packets, hence it cannot be used as indicative of the host being compromised. The number $F_2$, called the second frequency moment (also called the surprise number or self-join size) clearly indicates if the host has the characteristics of a compromised one. 

It is easy to see that the variance of the frequencies 
\begin{equation}
\sigma_F  = \frac{F_2}{F_0} - \left(\frac{F_1}{F_0}\right)^2
\end{equation}

Since in an \textsc{SDN}, the switches have very low memory and operational capacities, the space and computational complexity of maintaining the $F$ vector (e.g. $F_{0}$) for each host is prohibitive. Note that $F$ can be stored as a list of key-value pairs but since $|{\cal H}|$ is usually very large, even for small values of $M$ it will not be practical. Therefore, we cannot directly use $F_0$ nor $F_2$ because of space and computational problems.

To solve the computational and space problem, we use AMS sketch algorithm \cite{ams} to estimate $F_2$ for each host. Using the sketch for every host, the $F_2$ can be estimated in constant time per ingress packet and constant overall space. 

All switches use a set of $d$ random hash functions from a $4$-wise family of universal hash functions.

\begin{equation}
g_i : {\cal H} \rightarrow \{-1,1\}
\end{equation} 

Each switch uses these $d$ hash functions and maintains $d$ integer variables  ($X_{1}^{h}, X_{2}^{h}, \ldots, X_{d}^{h}$) for each host $h$. On arrival of ingress packets from the host $h$, it extracts the header $hdr$ from the packet and process as given in Algorithm \ref{algorithm_estimation_f}.

\begin{algorithm}[H]
	\caption{Algorithm for estimating $F_2$}
	\begin{algorithmic}
		\State $P \gets$ ingress packet from host $h$
		\State $hdr \gets \Call{ExtractHeader}{P}$
		\For{$i = 1$ to $d$}
		\State $X_{i}^h \leftarrow X_{i}^h + g_{i}(hdr)$ 
		\EndFor
	\end{algorithmic}
	\label{algorithm_estimation_f}
\end{algorithm}
After processing of the $M^{th}$ packet from host $h$, it returns the estimate for the value of $F_2$ for host $h$ as follows
\begin{equation}\label{mean_approximation}
F_{2}^{'} := {{\textsc{mean }}}(X_{1}^{2}, X_{2}^{2}, \ldots, X_{d}^{2}) 
\end{equation}

Equation \eqref{mean_approximation} returns the average value of the squares of the $d$ values saved for the host $h$. We use following results from \cite{ams} to argue that  $F_{2}^{'}$ is a very close estimate of $F_{2}$.

\begin{lemma}\label{expectation_Lemma}
	For $1 \leq i \leq d$, $E[X_{i}^{2}] = F_{2}$
\end{lemma}

Using the fact that the hash functions are from a $4$-wise universal family of hash functions, we get the following bound on the variance in the estimate.

\begin{lemma}\label{variance_Lemma}
	For $1 \leq i \leq d$, $\textsc{Var}(X_{i}^{2}) \leq 2F_{2}$ 
\end{lemma}
Combining Lemma \ref{expectation_Lemma} and Lemma \ref{variance_Lemma} we get the following results.

\begin{theorem}
	If $d = \frac{2}{\eps^{2} \delta} $, then     
	\begin{itemize}
		\item $Pr[|F_{2}^{'} - F_2| > \eps F_2] \leq \delta$. 
		\item Processing time per packet is constant (see Algorithm \ref{algorithm_estimation_f}).
		\item The space requirement is $O(\frac{1}{\eps^{2} \delta}) = O(1)$.
		
	\end{itemize}
\end{theorem} 

After approximating the $F_2$ (and respective integer variable $X$) for $d$ hash functions, we set a threshold $\tau$ to differentiate the zombie hosts from the good ones. If $X_{i}^{h} \leq \tau$, then we declare that host $h$ as the zombie host and vice versa. The threshold value $\tau$ is taken +/- $5$ from the actual value of total number of packets. This is so because we believe that the zombie host tends to request more unique headers (out of the total number of packets) as compared to the normal host. After categorizing each host as zombie or good, we compute true-positive and true-negative values by comparing our predicted hosts categories with actual ones.

\pgfplotsset{title style={at={(0.85,0.75)}}}
\pgfplotsset{every x tick label/.append style={font=\footnotesize}}
\pgfplotsset{every y tick label/.append style={font=\footnotesize}}
\pgfplotsset{compat=1.5}
\noindent
\begin{figure*}[h]
	\centering
	\footnotesize
	\begin{tikzpicture}
	\begin{axis}[title={},
	compat=newest,
	xlabel style={text width=3.5cm, align=center},
	xlabel={{\small $\tau$ \\ (a) $M$ = 50}},
	ylabel={Percentage},  ylabel shift={-5pt},
	xtick={1,2,3,4,5,6,7,8,9,10},
	xticklabels={45,46,47,48,49,50,51,52,53,54,55},
	height=0.75\columnwidth, width=0.78\columnwidth, grid=major,
	ymin=60, ymax=100,
	legend to name=commonlegend_mape_comparison
	]
	\addplot+[
	mark size=2.5pt,
	smooth,
	error bars/.cd,
	y fixed,
	y dir=both,
	y explicit
	] table [x={index}, y={tp}, col sep=comma] {Tau_50_numHashFuntions_4_numHosts_30_numSwitches_5_numHeaders_5000_numPackets_50.csv};
	\addplot+[
	mark size=2.5pt,
	dashed,
	error bars/.cd,
	y fixed,
	y dir=both,
	y explicit
	]
	table [x={index}, y={tn}, col sep=comma] {Tau_50_numHashFuntions_4_numHosts_30_numSwitches_5_numHeaders_5000_numPackets_50.csv};
	\end{axis}
	\end{tikzpicture}%
	\begin{tikzpicture}
	\begin{axis}[title={},
	compat=newest,
	xlabel style={text width=4cm, align=center},
	xlabel={{\small $\tau$ \\(b) $M$ = 100}},
	xtick={1,2,3,4,5,6,7,8,9,10},
	xticklabels={95,96,97,98,99,100,101,102,103,104,105},
	height=0.75\columnwidth, width=0.78\columnwidth, grid=major,
	ymin=60, ymax=100,
	legend to name=commonlegend_mape_comparison
	]
	\addplot+[
	mark size=2.5pt,
	smooth,
	error bars/.cd,
	y fixed,
	y dir=both,
	y explicit
	] table [x={index}, y={tp}, col sep=comma] {Tau_50_numHashFuntions_4_numHosts_30_numSwitches_5_numHeaders_5000_numPackets_50.csv};
	\addplot+[
	mark size=2.5pt,
	dashed,
	error bars/.cd,
	y fixed,
	y dir=both,
	y explicit
	]
	table [x={index}, y={tn}, col sep=comma] {Tau_50_numHashFuntions_4_numHosts_30_numSwitches_5_numHeaders_5000_numPackets_100.csv};
	\end{axis}    
	\end{tikzpicture}%
	\begin{tikzpicture}
	\begin{axis}[title={},
	compat=newest,
	xlabel style={text width=3.5cm, align=center},
	xlabel={{\small $\tau$ \\ (c) $M$ = 150}},
	xtick={1,2,3,4,5,6,7,8,9,10},
	xticklabels={145,146,147,148,149,150,151,152,153,154,155},
	height=0.75\columnwidth, width=0.78\columnwidth, grid=major,
	ymin=60, ymax=100,
	legend style={font=\footnotesize, column sep=1 ex, at={(1.55,0.75)},anchor=north east},
	legend columns = -1,
	legend entries={True-Positive, True-Negative},
	legend to name=commonlegend_mape_comparison,
	]
	\addplot+[
	mark size=2.5pt,
	smooth,
	error bars/.cd,
	y fixed,
	y dir=both,
	y explicit
	] table [x={index}, y={tp}, col sep=comma] {Tau_50_numHashFuntions_4_numHosts_30_numSwitches_5_numHeaders_5000_numPackets_150.csv};
	\addplot+[
	mark size=2.5pt,
	dashed,
	error bars/.cd,
	y fixed,
	y dir=both,
	y explicit
	]
	table [x={index}, y={tn}, col sep=comma] {Tau_50_numHashFuntions_4_numHosts_30_numSwitches_5_numHeaders_5000_numPackets_150.csv};    
	\end{axis}
	\end{tikzpicture}%
	\\    \ref{commonlegend_mape_comparison}    
	\caption{True-Positive and True-Negative values for increasing value of $\tau$ using $M$ as (a) $50$, (b) $100$, and (c) $150$ respectively. }
	\label{tp_tn_results}
\end{figure*}
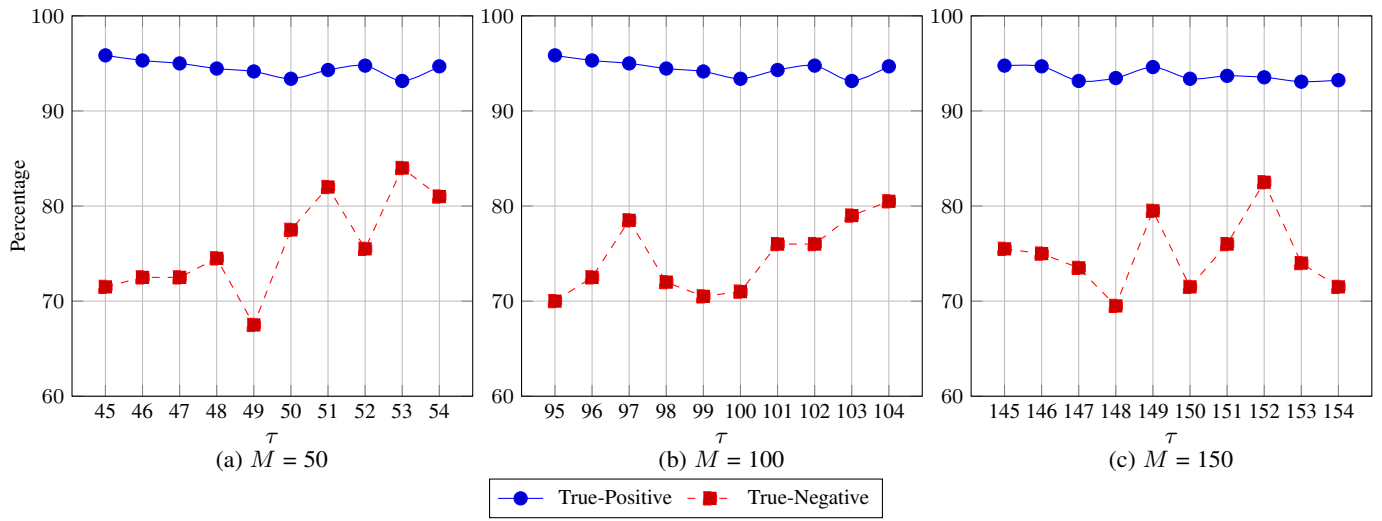

\section{Experimental Setup}\label{section_experimental_setup}
In this section, we describe the details of implementation and the hyperparameters list which we used for experiments. All results are computed by taking an average of $50$ runs.
Our code is implemented in Matlab on a Core i7 PC with 8GB memory, and 500GB storage. 
We report true-positive and true-negative values (for detecting the attackers and good hosts) as goodness measure.

First, we characterize the hosts as good/zombie hosts. A host is a {\em Good} host if it requests the destination addresses which are repeated multiple times (not mostly unique addresses). On the other hand, zombie host requests unique addresses most of the times, which cause DDoS attack. To detect an abnormal activity from a host $h$, we set a threshold $\tau$. If the approximate value for a host $h$ is less than $\tau$, it means that the host is a zombie and vice versa. We assume that each host can request to a single switch only and there are unique set of hosts connected to different switches. Each host can send a specific number of requests to the switch. Since we are approximating $F_{2}$, therefore we use several hash functions and take the mean of these approximations to compute as accurate results as possible. We fix the number of attackers that are randomly selected from a pool of hosts.

\subsection{Hyperparameters}
Our proposed method uses several hyperparameters (as discussed above) whose values combination effect the performance of our approach. Hyperparameters values yielding maximum performance are given in Table \ref{tbl2:hyper_parameters}.

\begin{table}[h!] 
	\centering
	
	\begin{tabular}{@{\extracolsep{4pt}}ll@{}}
		\hline
		Hyperparameters & Values 
		\\ 
		\hline \hline
		No. of Hosts & 30  \\
		No. of Hash Functions $d$ & 4 \\
		No. of Switches $N$ & 5 \\
		No. of Headers $\vert {\cal H} \vert$ & 5000 \\
		No. of Attackers & 4 \\
		No. of Packets $M$ & 50,100,150 \\
		Threshold $\tau$ & 1:10 \\
		\hline
		
	\end{tabular}
	\caption{Hyperparameters values.}
	\label{tbl2:hyper_parameters}
\end{table}

\subsection{Results and Discussion}
Results in Figure \ref{tp_tn_results} show the true-positive and true-negative values for the increasing value of $\tau$ by setting $M$ as (a) $50$, (b) $100$, and (c) $150$ respectively. 

We observe that the value of $\tau$ has not much effect on true-positive values if its value is taken almost same as the number of packets. This behavior shows that out of those total number of packets, if a host requests predominantly unique packets (headers), we assume that the host is not doing a normal operation. The same behavior is observed in case of true-negative (with just a few up/down spikes). The spikes in case of true-negative are because of the fact that in some cases, zombie host can try to act as normal host by varying its behavior. We tried to capture this effect also (by generating less unique requests) which sometimes cause the true-negative value to drop. However, adjusting the value of $\tau$ can eliminate this effect which is evident from Figure \ref{tp_tn_results}. If we further reduce the value of $\tau$, it means that we are giving more chances to zombie hosts to appear as normal hosts by slightly modifying their request pattern. On the other hand, if we increase the $\tau$, it means that we are forcing some of the normal hosts in the category of the zombie hosts which will reduce the efficiency of our system. Hence selecting the accurate value of $\tau$ is critical.

\begin{figure}[h]
	\centering
	\footnotesize
	\begin{tikzpicture}
	\begin{axis}[title={},
	compat=newest,
	xlabel style={text width=5.5cm, align=center},
	xlabel={{\small $d$}},
	ylabel={Percentage}, 
	ylabel shift={-5pt},
	legend style={font=\footnotesize, column sep=1 ex, at={(0.9,0.2)},anchor=north east},
	legend entries={True-Positive, True-Negative},
	height=0.75\columnwidth, width=0.78\columnwidth, grid=major,
	ymin=40, ymax=100,
	xmin=0, xmax=100,
	]
	\addplot+[
	mark size=1pt,
	smooth,
	error bars/.cd,
	y fixed,
	y dir=both,
	y explicit
	] table [x={index}, y={tp}, col sep=comma] {tp_tn_for_increasing_numHash_from_1_to_500.csv};
	\addplot+[
	mark size=1pt,
	smooth,
	error bars/.cd,
	y fixed,
	y dir=both,
	y explicit
	] table [x={index}, y={tn}, col sep=comma] {tp_tn_for_increasing_numHash_from_1_to_500.csv};
	\end{axis}
	\end{tikzpicture}%
	\caption{Effect of increasing $d$ on True-Positive and True-Negative values.}
	\label{accuracy_effect_on_increasing_hash_function}
\end{figure}
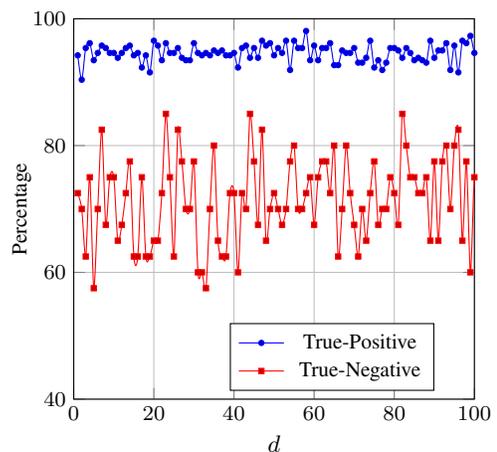

Results in Figure \ref{accuracy_effect_on_increasing_hash_function} show the effect of changing number of hash functions $d$ on true-positive and true-negative values. We can see that there are some up/down spikes but the overall trend does not change with the increasing value of $d$.

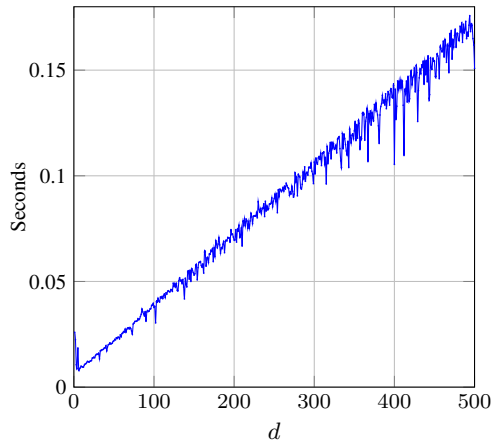
\begin{figure}[h]
	\centering
	\footnotesize
	\begin{tikzpicture}
	\begin{axis}[title={},
	compat=newest,
	xlabel style={text width=5.5cm, align=center},
	xlabel={{\small $d$}},
	ylabel={Seconds}, 
	ylabel shift={-4pt},
	ytick={0,0.05,0.1,0.15},
	yticklabels={0,0.05,0.1,0.15},
	xmin=0, xmax=500,
	height=0.75\columnwidth, width=0.78\columnwidth, grid=major,
	ymin=0, ymax=0.18,
	]
	\addplot+[
	mark size=0.1pt,
	smooth,
	error bars/.cd,
	y fixed,
	y dir=both,
	y explicit
	] table [x={index}, y={runtime}, col sep=comma] {runtime_for_increasing_numhash_functions_from_1_to_500.csv};
	\end{axis}
	\end{tikzpicture}%
	\caption{Effect of increasing $d$ on Runtime.}
	\label{runtime_effect_on_numHashFunc}
\end{figure}

Figure \ref{runtime_effect_on_numHashFunc} shows the effect of increasing value of $d$ on the runtime. We can see in Figure \ref{runtime_effect_on_numHashFunc} that the runtime almost linearly increase as we increase the number of hash functions. Therefore, the value of $d$ should be low.

\section{Conclusions}\label{conclusion}
In this paper, we proposed a time- and space-efficient solution to detect the DDoS attack on \textsc{SDN}. Most of the previously proposed methods require a large amount of data storage which is difficult to manage in case of memory limited devices. Our methodology distinguishes the attackers from the legitimate hosts by processing a stream of packet headers on the fly without consuming large space for data storage and processing. The conducted experiments prove the efficiency of our solution with a significantly good true-positive and true-negative rate. We also demonstrate that with an accurate value of $\tau$, we can achieve more accuracy in detecting the malicious attackers. 
One possible future extensions of this work is to investigate the better threshold values $\tau$ in comparison with the number of hosts.

\section*{Acknowledgment}
This research is partially supported by the Research Deanship of Islamic University of Madinah. 

\bibliographystyle{IEEEtran}
\bibliography{SDN_Open_Flow_Writeup}

\end{document}